\documentclass{article}
\usepackage{blindtext}
\usepackage{graphicx} 
\usepackage{multicol}

\usepackage{amsmath,graphicx,amssymb,amsthm, natbib}
\usepackage{lipsum}
\usepackage{microtype} 

\usepackage{authblk}

\setcitestyle{numbers}

\usepackage{url}

\usepackage{lineno}

\usepackage{xurl}

\setcitestyle{square}

\title{Measuring IT Carbon Footprint: What is the Current Status Actually?}
\author{Tom Kennes}
\affil{Skyworkz/SDIA, Amsterdam, The Netherlands}

\newcommand\keywords[1]{%
    \begingroup
    \let\and\\
    \par
    \noindent\textbf{Keywords:}\\#1\par
    \endgroup
}

\begin{document}

\maketitle

\begin{abstract}
Despite the new Corporate Sustainability Reporting Directive from the European Union, which presses large enterprises to be more transparent about their GHG emissions, and though large technology- or advisory firms might peddle otherwise, there are plenty of challenges ahead when it comes to measuring GHG emissions from IT activities in the first place. This paper categories those challenges into 4 categories, and explains the current status, shortcomings and potential future research directions. These categories are: measuring software energy consumption, server overhead energy consumption, Energy Mix and emissions from embodied carbon. Next to that, various non-profit and open-source initiatives are introduced as well as a mathematical framework, based on CPU consumption, that can act as a rule-of-thumb for quick and effortless assessments.

\end{abstract}

\begin{multicols}{2}

\section{Introduction}
\label{introduction}
The latest report by the IPCC \cite{IPCCReport2022} is clear. Urgent and drastic action is needed if we are to limit global warming to 1.5°C! The same holds true for the Information Technology industry (IT; the collection of software, digital applications, digital services and code running on servers). Even more so as the world accelerates its adoption of digital services and businesses increasingly tend to rely on software to drive their operations.

Notwithstanding, historically it has been difficult to put a number on the total energy consumed and GHG emission by IT worldwide, and recent research has been mostly dominated by figuring out appropriate assumptions and scope determination (\cite{andrae2015global}, \cite{belkhir2018assessing}, \cite{malmodin2018energy}, \cite{morley2018digitalisation}). This might explain why these discussions have not had more attention, a part of the public Netflix debate, to properly account for GHG emissions by IT. This debate, with the IEA and CarbonBrief (\cite{masanet2020recalibrating}, \cite{IEA2017digitalization}, \cite{CarbonBrief2020Factcheck}) on one side, and the Shift Project on the other side (\cite{ShiftProjectf2019towards}, \cite{ShiftProject2020did}), focuses on the environmental impact of streaming one hour of Netflix videos. Some of the disagreement can perhaps be best explained by Koomey's Law (\cite{koomey2009assessing}, \cite{koomey2010implications}), or the doubling of the energy efficiency of computers every 18 months leading to an increase in scientific awareness as to properly forecast the impact of future energy efficiency gains (\cite{andrae2019projecting}, \cite{andrae2020new}, \cite{koomey2021does}). All in all, the debate finely serves to illustrate some of the underlying issues.

Three years later and we still are not sure how accurate most of the top-down estimates really are. Hence, it might be more effective for the overall transition to focus on more accurate bottom-up measurements. Similar conclusions follow from \cite{worldwidewebOfCarbon} and \cite{mytton2022sources}. Top-down estimates might provide information for policymakers and global strategists, but they lock out workers that are responsible for and able to curb those same emissions. In the end, climate change is everybody's problem, so we should be involving everybody, as is argued by \cite{pasek2023world} as well.

As such, perhaps the numbers are accurate on some levels, but without lower-level trustworthy numbers it will remain difficult if not impossible to incorporate this into lower-level decision making and stimulate lower-level efforts. This is a problem, especially in areas where highly-skilled lower-level workers may have much more impact than higher-level global strategy management, such as IT.

For IT departments to lower their emissions, this implies empowering developers and engineers by tools that allow them to perform measurements, as well as stimulating management to incorporate this into their strategy. Luckily but perhaps unknowingly, IT management often already has tools and strategies for cost-management in place that lead to more sustainable IT practices. Lower IT utilisation is simply more green as well as more affordable. The resulting decrease in emissions, whether incidental or not, is something praiseworthy. Especially if the exact number could also be reported and audited. 

However, almost every IT department is either not yet measuring its GHG emissions, or putting out laudable goals with questionable details. And for good reason, it is simply not straightforward to measure the GHG emissions of software, code and IT infrastructure. Despite the multitude of engineering efforts to bridge this gap, there still exist significant problems that need tackling.

This paper focuses on those gaps. Coming from a practical point of view, measuring GHG emissions due to IT can be divided into 4 elements. These elements are:
\begin{itemize}
\item Software Energy Consumption ($E_{s}$)
\item Server Overhead Energy Consumption ($E_o$)
\item Momentaneous Energy Mix ($I$)
\item Embodied Carbon Emissions ($M$)
\end{itemize}

For each of those elements, this paper details the current status and possibilities, and proposes various research questions to advance our overall capability. The goal is to better align open-source initiatives, practitioners and academic research in the future. It is believed that, tackling these problems and improving our overall capability to measure the energy consumption and GHG emissions of software, will aid significantly in reducing the carbon footprint of IT.

Finally, note that these 4 elements need to be combined in order to calculate the actual carbon emissions. Logically, these can be separated into, on the one hand, operational carbon emissions, the energy consumed by the software and the overhead multiplied by the carbon emissions related to producing that energy in the first place, and, on the other hand, the embedded carbon emissions. In short:
\begin{equation}
    TotCarbon = \big((E_s + E_o) * I\big) + M
\end{equation}

The observant reader will recognize the formula as the Green Software Foundation (GSF) Software Carbon Intensity (SCI) specification (\cite{gsfsci}) at the moment of writing. The difference is the separation of Software and overhead, as well as the absence of what the SCI refers to as "the functional unit" (R):
\begin{equation}
    \frac{TotCarbon}{R} = SCI = \big((E * I\big) + M)    per R
\end{equation}

Although the GSF SCI is a good effort coming from renowned practitioners, by opting for flexibility, it does not dig deeper into the underlying problems. Hence, adoption remains difficult. At the same time, the SCI is getting traction and becoming a starting point for budding sustainable software engineers. It therefore remains a great tool for awareness creation and education, and a recommended read.

\section{Software Energy Consumption}
The most reliable method of measuring energy consumption by appliances remains using physical multimeters, whereas obtaining accurate empirical power measurements through software remains a challenge. According to a recent academic survey, focusing on the energy consumption of machine learning models \cite{garcia2019measure} there are roughly 2 ways to do so:
\begin{itemize}
    \item Power Monitoring Counters (PMC)-based models
    \item Simulation models based on computer-architecture
\end{itemize}
Note that out of the 19 analysed methods, only 6 are not based on PMC.

In this paper, we offer a third simplified way in order to both bring together our knowledge to more expressive mathematical formulas as well as to communicate and lobby the audience where to direct focus.

\subsection{PMC-based measurements}
Examples of PMCs commonly reflect statistics related to CPU instructions, memory- or disk-operations and resource stalls. In the past, they were mostly relevant for hardware development and testing, like for example chip performance and chip heat dissipation. As such, there has been ample research on the topic of PMCs, see for example \cite{goel2012techniques}, including the development of various models, most notably RAPL (Running Average Power Limit) \cite{RAPL2014}. Another popular model is IPMI, but it is often considered to be insufficiently accurate (\cite{2016IPMIAccuracy}).

RAPL has been developed by Intel, and is said to have reasonable accuracy (\cite{khan2018rapl}, \cite{hackenberg2015}, \cite{MeasuringEnergyConsumption2012}, \cite{ValidationDRAMRAPL2016}), although there has also been research advocating more cautiousness \cite{paniego2018analysis}. All in all, the true accuracy remains a bit difficult to determine and seems to be highly affected by different computer architectures, BIOS settings and further performance configurations, which makes adoption difficult. These are merely observations from practice as there has been a lack of research on the effect of these factors. There are some efforts to come to a generalised model based on RAPL data, which are discussed in the next subsection.

As an important sidenote, RAPL has a known vulnerability and corresponding mitigations (\cite{RAPLVULN2022}).

\subsection{Existing Data and Tools}
Accurate or not, at the moment of writing there already exist numerous energy measurement tools and applications, that are either directly powered by RAPL (see, among others, \cite{kavanagh2019rapid}, \cite{scaphandre}, \cite{joularjx}, \cite{kepler}, \cite{aiPowerMeter}, \cite{codeCarbon}, \cite{powerstat}, \cite{powertop}, \cite{intelPowerGadget}, \cite{GreenMetricsToolGreenCodingBerlin}) or using forecasts based on earlier collected measurements of (somewhat) identical machines (see, among others, \cite{SpecPowerModelGreenCodingBerlin}, \cite{rteil2022interact}, \cite{teadsassessments2}, \cite{CloudCarbonFootprint}). These earlier collected measurements are generally either obtained via SPEC (Standard Performance Evaluation Corporation) \cite{SpecPower} or collected through RAPL in fully controlled environments (\cite{teadsassessments1}, \cite{teadsassessments2}), or even both.

SPEC has been executing performance benchmark assessments for over 30 years, mostly through direct power measurements. Although there are indications that these assessments are reasonably rigorous (\cite{specDescriptions}, \cite{SpecPower}, \cite{specOverview}), it is known that replication of these measurements is a challenge, in particular due to specific performance-enhancing BIOS settings. Perhaps this explains the lack of audits of the SPEC benchmark assessments. Next to that, it is not clear to what extent the assessments can be generalised, nor whether they are nowadays still relevant to assess the performance and energy consumption of modern cloud environments, which have changed significantly over the last decade. For example, the latest benchmark (the SPEC Cloud IaaS assessment of 2018 \cite{specCloud}) attempts to measure generic Cloud IaaS performance using two specific exercises:
\begin{itemize}
    \item Stress Testing Apache Cassandra, using the Yahoo Cloud Serving Benchmark (see also \cite{cooper2010benchmarking})
    \item K-means clustering using Apache Hadoop (see also: \cite{issa2016performance})
\end{itemize}
It should be clear that these Cloud IaaS assessments only apply to a very narrow context and thus do not allow drawing conclusions about overall Cloud or Cloud IaaS performance. Cloud consists of an extremely vast set of services and configurations that are continuously evolving at many different levels. Hence, it is not useful to assess Cloud as a single entity without assessing its underlying services and components.

All in all, the effort by SPEC to develop these benchmarks and come up with measurements that can in turn be used to assess the energy consumption during runtime is praiseworthy. It's a great initiative to come to a centralised database, yet the need for more rigorous measurements, transparency, replicability and academic research should be clear.

\subsection{Potential Improvements and Research Questions}
Given the analysis in the previous section, there are a couple of research questions and directions:
\begin{itemize}
    \item Continue the RAPL assessments and development. From a practical point of view, the requirements for root access on a machine significantly complicates adoption. If this could be alleviated, this should be more straightforward.
    \item Improve our understanding of BIOS and performance configuration settings and their actual impact on performance vs energy consumption. 
    \item Further integration of RAPL energy assessments with applications, software and code
    \item Continue the SPEC initiative, but in a more transparent and rigorous manner. The focus should lie as much as possible on lower level components, rather than the specific software that is actually causing the energy consumption. The direct relationship between CPU and energy consumption is much more interesting than the indirect relationship between software and energy consumption. Besides, CPU and memory usage is easier tracked in most environments. Yet, the latter might be helpful when deciding on IT architecture and costs.
\end{itemize}

\subsection{SDIA DEF Model}
It is believed by the SDIA that it is beneficial to have a rule of thumb for quick decision making and more rough estimations in environments where direct measuring is simply not possible yet (Cloud). This is why we introduce the SDIA DEF Model (Sustainable Digital Infrastructure Alliance Digital Environmental Footprint), which can also be used as a general mathematical framework to embed future research and development. The rationale behind the model involves bringing down somewhat complicated energy models into metrics that are easier to observe to some extent. In other words, CPU usage.

\subsubsection{Generic Formulations}
First of all, we split out a server into 4 of its core functionalities: CPU, memory, storage or IO and networking.

Given that the total energy consumed by a server can be written as the sum of the energy consumed by its components, as well as a baseline energy consumption representing idle mode, we can write:
\begin{equation}
    E_{tot} = E_{cpu} + E_{mem} + E_{IO} + E_{net} + \beta_{idle}
\end{equation}

Now, we can generally not measure the energy consumption for any of these components directly. However, we can split them into a factor we can measure, component usage ($U$), and a factor that translates that observable into energy ($f$). Note that we thus gather the underlying complexity of converting usage into energy into the second factor ($f$).

Any improvements in our ability to accurately measure software energy consumption would be by improving that second factor ($f$).

The total energy used by a server is thus defined as:
\begin{multline}
    E_{tot} = U_{cpu} f_{cpu}(U_{cpu}) + U_{mem}f_{mem}(U_{mem}) + \\ U_{IO}f_{IO}(U_{IO}) + U_{net}f_{net}(U_{net}) + \beta_{idle}
\end{multline}
Whereas:
\begin{itemize}
    \item $E_{tot}$: The total energy used by a server
    \item $PUE$: Power Usage Effectiveness. The overhead, mostly due to cooling.
    \item $U_{cpu}$: Current CPU usage, measured in number of CPUs
    \item $f_{cpu}(U_{cpu})$: Function of the CPU energy consumption per CPU usage in number of CPUs
    \item $U_{mem}$: Current memory usage, measured in bits
    \item $f_{mem}(U_{mem})$: Function of the memory energy consumption per memory usage in bits
    \item $U_{IO}$: Current storage usage, measured in bits
    \item $f_{IO}(U_{IO})$: Function of the storage energy consumption per storage usage in bits
    \item $U_{net}$: Current network usage, measured in bits
    \item $f_{net}(U_{net})$	: Function of the network energy consumption per network usage in bits
    \item $\beta_{idle}$: Energy consumption of the server when idling
\end{itemize}

Note that the functions denoted by $f_x(U_x)$ represent the energy allocation per unit of usage. Its unit is thus $E/U_x$. 

Also note that these are not necessarily linear. The functional depiction does not assume anything about the nature of the relationship between component usage and energy consumption. There might be other factors that should also be incorporated, such as server type or server age. 

\subsubsection{Assumptions}
The Thermal Design Power (TDP), or the maximum amount of heat a CPU is designed to be able to dissipate without overheating, and is thus often used as a proxy for the energy consumed by a CPU when it runs on max performance, see for example \cite{sun2019summarizing}. Nonetheless, we admit that TDP may not be the best metric to evaluate power consumption. Chips can easily consume more power than their TDP specification for a short period of time. However, TDP can be a
reasonable estimation of the average power consumption when a CPU runs regular workloads at its base clock.

\textbf{Assumption 1}
We can use the TDP as a proxy for energy consumption by a single CPU core when fully using the CPU

Because the energy consumption of the other components generally is more difficult to predict, we relate those directly to the energy consumed by the CPU in the form of fixed allocation factors ($\mathbf{\alpha}$).

\textbf{Assumption 2}
For server running on full performance:
\begin{itemize}
    \item $\alpha_{CPU}$ of the energy consumed by the server is due to CPU
    \item $\alpha_{mem}$ of the energy consumed by the server is due to memory
    \item $\alpha_{IO}$ of the energy consumed by the server is due to storage
    \item $\alpha_{net}$ of the energy consumed by the server is due to networking
    \item All energy of a server can be allocated to these 4 components: $\alpha_{CPU} + \alpha_{mem}+\alpha_{IO}+\alpha_{net} = 1$
\end{itemize}

More often than not, however, we are not running a server on full performance. In order to estimate the energy consumed in those scenarios, the SDIA proposes to use linearity.

\textbf{Assumption 3}
The energy consumed by a server increases linearly relative to the increase in usage of any of its components.

\textbf{Assumption 4}
Idling consumption is expected to be zero. After all, if none of the components of the system are using energy, the system itself should logically also not be using energy.

\subsubsection{Derivation}
Let $E_{tot-max}$ denote the total energy consumed by a server when running on full performance.

And, using the first assumption:
\begin{multline}
    E_{cpu-max} = \\ U_{cpu-max} f_{cpu}(U_{cpu-max}) = TDP * N_{CPU}
\end{multline}

Whereas:
\begin{itemize}
    \item $N_{CPU}$ denotes the number of CPUs
\end{itemize}

Then, using the second assumption:
\begin{equation}
    E_{tot-max} = E_{cpu-max} /\alpha_{CPU}= [TDP * N_{CPU}]/\alpha_{CPU}
\end{equation}

Substituting $E_{cpu-max}$ by $U_{cpu-max}f_{cpu}(U_{cpu})$ , using our third assumption, and solving for $f_{cpu}(U_{cpu})$, we get:
\begin{equation}
    f_{cpu}(U_{cpu}) =( TDP * N_{CPU})/U_{cpu-max}
\end{equation}

And finally for the energy related to the CPU:
\begin{multline}
    E_{cpu} = U_{cpu}f_{cpu}(U_{cpu}) = \\ (U_{cpu}/U_{cpu-max})( [TDP * N_{CPU}])
\end{multline}

For the other components, we get:
\begin{multline}
    E_{tot-max} = E_{mem-max}/\alpha_{mem} = \\ [U_{mem-max}f_{mem}(U_{mem-max})]/\alpha_{mem}
\end{multline}

Plugging in for $E_{tot-max}$, and solving for $U_{mem}f_{mem}(U_{mem})$:
\begin{multline}
    \alpha_{mem}[[TDP * NCPU]/\alpha_{CPU}] = \\ U_{mem}f_{mem}(U_{mem})
\end{multline}

Or:
\begin{multline}
    (\alpha_{mem}/\alpha_{CPU})[TDP * N_{CPU} ] = \\ U_{mem-max}f_{mem}(U_{mem-max})
\end{multline}

Solving for $f_{mem}(U_{mem})$, and since $f_{mem}(U_{mem})$ does not actually depend on $U_{mem}$:
\begin{multline}
    f_{mem}(U_{mem}) = \\ (\alpha_{mem}/\alpha_{CPU})[TDP * N_{CPU}] / U_{mem-max}
\end{multline}

Meaning we end up with the total energy related to memory:
\begin{multline}
    E_{mem} = \\ (U_{mem} / U_{mem-max})(\alpha_{mem}/\alpha_{CPU})[TDP * N_{CPU}] 
\end{multline}

The energy related to the other components can then be found in a similar fashion, due to symmetry.

Given that the allocation vector \textbf{$\alpha$} is a constant, the energy functions $f_x(U_x)$ become constants in this case as well. This implies that the partial derivatives, the marginal energy cost related to usage of component $x$, can be simplified to that constant. Which gives us back our third assumption:
\begin{equation}
    \partial E_x / \partial U_x =  f_x(U_x)
\end{equation}

\subsection{Potential Improvements and Research Questions}
\subsubsection{Empirical Estimations Of CPU Energy Usage}
Note that these estimations do away with a logical formulation of $f_{cpu}(U_{cpu})$, and instead empirically measure the energy consumption when running on a certain CPU load. This data, collected through experimentation, can then serve as input for statistical models to estimate $f_{cpu}(U_{cpu}$). 

\subsubsection{Empirical Estimations Of Other Component Usage}
Whereas there exist thus several initiatives that attempt to make sense of the CPU energy consumption, I am not aware of any parties that do so for the other components. Perhaps it is also not as relevant as CPU?

\subsubsection{Empirical Validation of Assumptions}
Although the assumptions laid out in the SDIA methodology are limiting, they still provide means to obtain a number for energy consumption by a server. Furthermore, it is not clear how inaccurate these numbers are, and if scenarios exist where these numbers are valid nonetheless. These research questions tie directly into the assumptions:
\begin{itemize}
    \item How realistic are the numbers for the allocation factors in the example?
    \item Does linearity make sense? If so, in which cases? Which settings, which CPUs, which architectures, or maybe even which kind of applications?
    \item Can TDP actually be used as a proxy for CPU energy consumption at full performance?
\end{itemize}

\subsubsection{Empirical Validation of Idle Consumption}
The last assumption has not been given enough attention in the formulas, but it might be a swift assumption that systems without energy consumption by CPU, memory, IO and networking can still be using energy. The system of course has to maintain an operating modus, but it should be logical that this energy consumption is negligible compared to the energy consumed by CPUs when they are in usage.

Nonetheless, a measure of idle consumption might be extremely relevant in the decision to lower load. If consumption when idling is relatively high, powering off complete machines, or nodes would be a preferred strategy to lower energy and emissions, rather than simply distributing and lowering the load on each machine within a cluster.

\begin{itemize}
\item When none of the components are consuming energy, what is the idling consumption of a server?
\item Should idling consumption be incorporated into the formulas? If so, how?
\end{itemize}

\section{Overhead: PUE}
\subsection{PUE, and Other Available Metrics}
From an energy perspective, running something on a server is mostly a conversion from electrical energy to thermal energy. Commonly, fans are used to divert the heat away in most devices, but more sophisticated infrastructure, like modern data centres., might use more advanced solutions, like  liquid cooling. No matter which form of equipment is being used, these necessary appliances and components require energy to run as well, and therefore can be regarded as overhead from a software perspective. 

For data centres., the energy used by these overhead physical components is most often reported as the Power Usage Effectiveness (PUE), which compares the overall electrical energy used by a datacenter to the electrical energy that actually is consumed by the hardware. Note thus that the measure is commonly used when assessing overhead within data centres., but the same idea could be applied on more individual devices.

Over time, the PUE metric has become the de-facto standard, and has also gone some way to creating competition, driving efficiencies up as advertised PUE values become lower. Standards for measurement and scope-determination have been set up and most companies, especially hyperscalers boasting efficient PUE numbers, provide somewhat detailed reports on the topic. For more information, see the following whitepaper from the Green Grid Foundation \cite{greengrid_PUE}. 

Mathematically, you would get:
\begin{equation}
E_{tot-cpu} = E_{cpu}*PUE
\end{equation}

Nonetheless, the metric has since then been under heavy scrutiny. As argued by \cite{brady2013case}, PUE has some known issues:
\begin{itemize}
    \item PUE does not include the energy efficiency of the IT equipment
    \item PUE does not reflect meteorological conditions of the facility. In colder climates less energy might be needed for cooling purposes. Note that this makes a crude comparison based only on PUE without climate reporting somewhat meaningless.
    \item The PUE metric is more useful for data centres that are operating at full IT capacity, as this should mean that the supporting infrastructure is also operating at its designed capacity
    \item PUE does not include waste heat recycling
\end{itemize}

This is also why industry experts and researchers have been advocating for the usage of additional metrics more focused on capturing a broader sense of energy performance (see \cite{salim2020green}, \cite{van2017analysis}), while also pointing out that efforts to improve the energy efficiency of the physical infrastructure of data centres only lead to marginal improvements, and that focus should be on IT instead \cite{bashroush2020beyond}. These metrics include:
\begin{itemize}
    \item CUE: Carbon Usage Effectiveness
    \item WUE: Water Usage Effectiveness
    \item ERF: Energy Re-Usage Factor
    \item ITEE: IT Equipment Efficiency
    \item OEM: On-site Energy Matching (which focuses on on-site renewables)
\end{itemize}

Although data centres have become much more efficient over the past decades in relation to all of these metrics (\cite{koomey2009assessing}, \cite{koomey2010implications}), and software is expected to quickly become more carbon aware in the future, there remains an apparent division between the two disciplines as shown by these metrics. In other words, these metrics mostly serve to properly communicate efforts of data centres to become more sustainable over longer periods of time, instead of informing software developers of the instantaneous waste by the underlying infrastructure when running their code.

Partially, this can be explained by the creation of abstraction layers between the software and the hardware. But there are also commercial reasons behind limiting the transparency around one's operational efficiency. No matter the reason, in the end we are all in the same boat and there are already companies that intend to capitalise on this divide by offering a combination of hardware, software, reporting and transparency. Next to that, the data centre design space is rapidly innovating in several directions. Heat reusage, better cooling and hardware custom designed for specific use-cases are expected to become more and more prevalent in the future.

\subsection{Potential Improvements and Research Questions}
The biggest challenge lies in getting numbers out of data centres into the heads of the consumers of those data centres. E.g. developers, engineers, managers and the companies they work for. The solution not only lies in more transparency, but might also require changes in data centre design. Not all data centres allow for an effective measurement of the energy consumption of individual software, and hence might be forced to turn to aggregates.

Next to that, although PUE aids in obtaining better estimates on energy consumption of software, since it is generally aggregated over a year it makes it impossible to get accurate instantaneous numbers.

Finally, the mathematics in this section have focused on PUE only, yet, with new waste heat recovery methods finding their ways in modern data centres., it might be better to account for this. Perhaps in a single metric representing the amount of overhead or unrecoverable resources after running a certain workload in some datacenter.

\section{Momentaneous Energy Mix}
The momentaneous Energy Mix represents the combination of energy sources currently used to satisfy our society in its hunger for energy. In contrast to PUE, we do have somewhat trustworthy numbers on the energy production side. We can fairly trustworthy measure the amount of Energy generated by renewable sources vs burning fossil fuels, and in turn the GHG emissions of the current Energy Mix. Although there are some sources available (like ElectricityMaps \cite{electricitymaps} and Wattime \cite{wattime}), and even tools to easily retrieve data from these sources (\cite{carbonAwareSDK}, \cite{kubernetescarbonintensityexporter}, \cite{gridintensitygo}), it is more challenging to properly connect to the consumption of this energy. There is a discrepancy in time granularity between energy production and energy consumption. That is, if it is only possible to get reliable numbers on the emission of energy production per hour, you would also need to aggregate the software energy consumption per hour in order to perform calculations using the two variables. There is thus somewhat of a gap between theory and practice here, as we would like to directly compare the electricity Energy Mix of a particular moment with the consumption of that energy in that particular moment. 

In order to again provide a generic model and then introduce physical and practical limitations we first integrate over time to later simplify to a Riemann sum over the smallest granularity of the two measurements. Mathematically, we would ideally get:
\begin{equation}
kgCO2 =PUE * \int_0^TGM(t)*E_{tot}(t)dt
\end{equation}
Wheres:
\begin{itemize}
    \item $T$ represents the total amount of time $GM(t)$ the Energy Mix at time $t$
    \item $E_{tot-cpu}(t)$ the total energy consumption by the server	
\end{itemize}

But, given that we can only get 30-minute Energy Mix updates:
\begin{equation}
    kgCO2=PUE*\sum_{t_i = 1}^TGM(t_i)\int_{=ti-1}^{t_i}E_{tot}(t_j)
\end{equation}

The same can be said for the fact that we do not continuously measure energy consumption, Hence:
\begin{equation}
    kgCO2=PUE*\sum_{t_i = 1}^TGM(t_i)\sum_{\tau=ti-1}^{t_i}E_{tot}(\tau)
\end{equation}
Notice thus that we might lose a bit of accuracy by smoothing out over time. Also note that we focus here on CO2 emissions. 

Given that we know the GHG emissions related to generating energy from both renewables and non-renewables, we can get numbers for these other GHGs by symmetry as well.

Finally, note that we generally focus on Energy (Joules) directly, but these formulas can also be extended to either cover watts or Kilowatt-hours.

\subsection{Potential Improvements and Research Questions}
The discrepancy in time-granularity remains an issue, which we generally only can solve by creating aggregations over time. Though, the same remark can be made with regard to measuring Software Energy Consumption through RAPL, since these tend to consist of more instantaneous snapshots rather than continuous measurements. 

Hence:
\begin{itemize}
    \item How reliable are Software Energy Measurement aggregations over time given that their direct measurements can miss momentaneous peaks? Should we do better? And can we do better?
    \item How reliable is aggregating PUE over time? Should we do better?
    \item How reliable is aggregating the Energy Mix over time? Should we do better? And can we do better?
\end{itemize}

\subsection{Average vs Marginal}
Furthermore, there is ongoing debate between the usage of marginal CO2 emissions versus the average CO2 emissions when calculating the resulting CO2 emissions from having consumed some amount of energy. Note that this is a completely different scientific field of its own, with great importance. After all, modelling the modern energy grid with all its producers and consumers is a complicated task. However, it is extremely important to accurately assess the impact of globally changing consumption patterns due to, for example, demand shifting and government policies. 

Marginal emissions reflect the additional emissions coming from generating the additional required energy on top of some baseline energy requirements. In contrast, average emissions reflect the average emissions of all energy producers delivering energy to the current Energy Mix.

In theory, as argued by \cite{schram2019use}, reducing the energy consumption would be to reduce the energy production at those facilities that are operating at the margin, or those facilities that are operating at the highest weighted combination of fuel costs, emissions and operational costs. Therefore, it would make sense to relate marginal energy demand to marginal energy supply. In academic literature, the result of a change in systemic energy demand has been shown to be more accurately measurable when using marginal emissions. Since it is rather difficult to simulate the entire grid with all its producers and consumers, these experiments are solely based on rather tightly controlled simulations, like \cite{hittingerEmissionFactors}.

Despite the historic backing for marginal emissions over average emissions by many academics, recently, (academic) criticism has increased. As argued by \cite{aukehoestra} and \cite{marginalvsaverage}, among others, stipulating which appliances and activities are part of the marginal energy demand, and thereby establishing a order of merit for energy consumption, is near impossible. In other words, should the emissions of the fossil-fuel based power plant be attributed to one extra EV charging cycle, your fridge or your neighbors jacuzzi? Stipulating the energy demand of one of these appliances as marginal, would allocate all marginal emissions, which often are fossil-fuel-based, to that appliance thereby marking the energy consumption of the appliance as unnecessary or even polluting. 

In the inverse scenario, one where all energy demand is met by renewable energy and marginal demand is to be met by burning fossil fuels, it is more tempting to again conclude that marginal demand is met by marginal supply. Yet, the same argument holds; it is impossible to establish an order of merit, whether demand is new or existing.

Perhaps it is best to be pragmatic in this discussion. The final goal is to reduce GHG emissions, and that might imply looking at averages in most scenarios.

\subsection{Power Purchasing Agreements and Guarantees of Origin}
However, for the sake of transparency it might in some scenarios be best to make use of averages, especially when dealing with the impact of Power Purchasing Agreements between a large consumer and producer that come with Guarantees of Origin (GO). In essence, these contracts form an agreement of future delivery of energy against a certain price, whereas that energy is up to some extent coming from renewable sources. Note that this does not directly mean that the generated electrons that actually arrive at the consumer, have been generated by those renewable sources. We are simply unable to track electrons over the vast internationally interconnected grid, despite what some companies might want you to believe \cite{irishcompaniesdropmisleading}. In other words, without direct connections to a renewable energy source, it is only possible to claim you are running on 100\% renewables if the actual renewable energy source is located on-premise.

Nonetheless, for energy producers it is also interesting to offer these agreements in order to invest in the production of these renewable energy sources, even though the GOs are by some academics regarded as ineffective policies \cite{IsguaranteeoforiginreallyaneffectiveenergypolicytoolinEuropeAcriticalapproach}. This increase in the production of energy by renewable sources can lower the average emissions over time of a region as a whole, rather than just merely the purchaser of those GOs. If and only if some of the non-renewable production can be replaced.

\section{Embodied Carbon}
The scientific field of calculating Embodied Carbon can be regarded as a field on its own, more rooted into Life Cycle Assessments (LCA) than runtime (operational) measurements. In short, Embodied Carbon involves attributing the emitted carbon (or any other GHG emissions for that matter) related to creating/manufacturing, transport, maintenance and disposal of the resource, to the user of that resource. In a sense, you thus make the consumer of that resource responsible for the emissions that were necessary to make and keep that resource consumable in the first place, including proper disposal afterwards.

These resources generally can take up a broad definition, but within the domain of IT and data centres, we often talk about the hardware (server racks), cabling and cooling networks, as well as the overhead physical housing. 

This also explains why measuring the Embodied Carbon asks for a completely different approach. Fortunately, the theory is  straightforward. We only need to get these manufacturing-, transport- and recycling/disposal-related emissions for each individual physical piece of equipment involved, and attribute the amount of time it was used by a certain process proportional to the total active lifetime of this individual physical piece of equipment. Mathematically, you would thus sum over the total number of objects (n) and their involved Embodied Carbon:
\begin{equation}
    kgCO2=\sum_{i= 1}^n\left[(M_{o_i}+R_{o_i}+EoL_{o_i})(t_{c,o_i}/T_{o_i})\right]
\end{equation}

Whereas:
\begin{itemize}
    \item $n$: the total number of objects
    \item $o_i$: a specific object, indexed as object $i$
    \item $M_{o_i}$: The carbon emission involved during manufacturing of object $i$
    \item $R_{o_i}$: The carbon emission involved during repairments/maintenance of object $i$
    \item $EoL_{o_i}$: The carbon emission involved during End-of-Life disposal of object $i$
    \item $t_{c,o_i}$: The total time some consumer $c$ consumed object $i$
    \item $T_{o_i}$: The total time object $i$ remains in a consumable state (usage phase lifespan)
\end{itemize}

In theory, these calculations are thus relatively straightforward. In practice, a number of difficulties are likely to arise: 
\begin{itemize}
    \item There might be thousands if not millions of objects involved, which all have their own life-cycle related statistics.
    \item These objects have different, potentially overlapping, usage phase lifespans $T_{o_i}$.
    \item Many resources can be shared by multiple consumers at the same time. Servers are shared by multiple processes, racks by multiple servers, buildings and cables by multiple racks, etc.
    \item Transparency and auditability of the embedded carbon of these objects. It can sometimes be difficult to get these numbers, if trustworthy at all.
\end{itemize}

Next to these difficulties, there are also a couple of problems likely to arise, which might require a more holistic, perhaps even legislative, solution: 
\begin{itemize}
    \item Scope definitions should be extremely clear for these calculations to give meaningful results. For many of the life cycle stages, these definitions are not very strict and thus lead to reporters taking advantage of the consequential leeway.
    \item Who is responsible for the embodied carbon when an object is in the Usage-phase, but actually not being used? E.g. when there actually is no consumer to attribute to.
\end{itemize}

\subsection{Life Cycle Statistics}
For many products, especially those used in construction and physical engineering, we already require manufacturers to disclose the carbon emitted during manufacturing as well as the expected emissions during recycling or waste disposal. 

Although manufacturers might have an incentive to underestimate these carbon emissions, especially when they are getting penalised for their emissions, these should in general be relatively trustworthy. In the end, it is a somewhat replicable physical process, so in theory it should well be possible to audit these numbers.

Additionally, for objects with long life cycles it might be difficult to predict how much carbon will be emitted during recycling or disposal. Who knows where our recycling technology will be in 10 years, let alone 50 years or more?

\subsection{Different Lifespans}
Depending on how far you are willing to go to get more accurate numbers, you might need to account for replacing objects or parts that have shorter lifespans, including back-up systems when these objects are part of a continuous service without downtime. However, when you would be the consumer of some object oi and its replacement part $o_j$, you do not have to consider how these objects were related in the first place. You merely care how long you have been a consumer of these objects.

\subsection{Sharing of Objects and Resources}
Properly attributing embodied carbon to individual consumers can already be quite cumbersome when only dealing with the time- and object- dimensions. However, ideally you also would want to properly attribute the sharing of resources. If two consumers can make use of the same resource at the same time, without a decrease in quality of service generated by that resource, that resource would be used much more efficiently. So, evidently, this happens a lot within data centres and it is for good reason data centre operators are eager to see technologies like hypervisors and virtualization being developed further to further push the efficiency (and in turn profitability) due to sharing of resources. Hence, sharing of resources is quite common and it is likely it will only be more common in the future.

However, proper carbon accounting and attributing embodied carbon downwards becomes more difficult, especially as the sharing model becomes more complex as well. In some moments in time, you might have an $x$ number of shared consumers, and for other moments it might be different. Hence, if you intend to properly account for the sharing of resources over time, you would need to calculate this for each unit of time available. 

Thus, for example, in order to accurately attribute the embodied carbon of a building to one of the consumers that consumes one of the servers housed within that building for an hour, you would need to know: The lifespan of the building as a whole And how much of the total capacity is being used by that consumer for each unit of time. Be it per second, or per hour, depending on how much this would shift over time.

Mathematically, for each consumer you would get:
\begin{multline}
    kgCO2 = \\ \sum_{i=1}^n\left[(M_{o_i}+R_{o_i}+EoL_{o_i})(1/T_{o_i})\sum_{\tau=1}^{t_{c, o_i}}f_c({\tau, o_i})\right]
\end{multline}
Whereas:
\begin{itemize}
    \item $\tau$: denotes the individual steps in time between 1 and $t_{c, o_i}$, the total time user u consumed resource $o_i$
    \item $f_c({\tau}, {o_i})$: denotes how much of resource oi is being consumed by consumer $c$ at time . Note that this is a number between 0 and 1, a $1$ indicating that there is no sharing of resources whereas, for example, a $0.5$ would indicate that 50\% of the resource is being consumed by consumer $c$, and the other 50\% by other consumers.
\end{itemize}

Hence, accounting for dynamic sharing of resources over time can greatly complicate the calculations, and in turn introduce a lot of data requirements. Both from a practical and theoretical point of view, there are objections to the necessity of going this far. It might for example not even be feasible from a practical point of view to know how many consumers of a certain resource exist for each moment in time, or having to keep track of it might introduce too much overhead. However, if you are working with reservations, you would be able to better predict the sharing factors beforehand and you would thus not need to track this actively over time.

From a theoretical point of view it is also questionable whether it would be necessary in the first place. If the total number of consumers of a resource is stable over time, as well as their sharing factors, you might be able to get reasonably accurate numbers by the factors out over time.

\subsection{Scope Definitions and Idle Time Usage}
Whereas the difficulties are expected to be surmountable when reporting on Embodied emissions, without proper legislative guidelines it might be much more difficult to overcome the problem of scope and allocating Idle Time within the usage phase.

Whereas scope definitions have been debated intensively over the last decade, discussions around the attribution of embodied carbon during idle time usage are just starting. And it is destined to be an interesting field filled with controversy as well, with a lot of different use cases both within the IT and data centres as well as the bigger public domain. 

For example, take the football World Cup of 2022 hosted by Qatar organised by FIFA. Qatar has built several new stadiums to host the matches, which are expected to have a lifespan of only 60 years. Although there are some plans here and there to reuse these stadiums for other events in the future, it is unlikely these stadiums will be filled most of the time. FIFA, in their Carbon Accounting report, has published the embodied carbon related to the stadiums and the event as a whole. In the report, they only take responsibility for the 26 days during which the World Cup unfolded. The other days within the 60 year lifespan are not taken into account. From a carbon accounting perspective, as well as the formulas in the previous sections, they are right. However, it remains suspicious given that these stadiums were built specifically for the event. In this particular, who should be held responsible for those days when the stadiums are not being used within their lifespan? FIFA? Qatar? Football supporters visiting the event?

Of course, there must also be venues in the US and in Europe for which the same can be said. Climate change is a global problem. Proper accounting of embodied carbon as well as taking up responsibility for it, have global impact as well.

\subsection{Potential Improvements and Research Questions}
\begin{itemize}
\item As mentioned before, most of the work when calculating embodied carbon emissions involves executing or obtaining proper Life Cycle Assessments (LCA). In short, how reliable are the constants in the equations ($M_{o_i}$, $R_{o_i}$, $EoL_{o_i}$, $T_{o_i}$).

\item If these are reliable, it is merely a matter of averaging out where possible. The question then becomes for which scenarios this would be harmful and when you are better off doing so. When can we average things out over time? How dynamic are these resource consumption factors over time?

\item Next to that, which objects should be within scope? How can we further define this properly such that everybody can start reporting adequately and consistently?

\item Finally, who should be held responsible for idle time usage embodied carbon emissions?
\end{itemize}

\subsection{Note}
Although this section has mostly focused on embedded carbon emissions, by symmetry you could use the same equations for embedded emissions related to other GHGs.

\section{Conclusion}
The field of sustainable IT has heavily been progressing during the time of research and writing for this article, and it is hoped that this trend will only accelerate in the future. Even though there are many different stakeholders with different incentives, we are optimistic about the future, especially because people show an overwhelming willingness to incorporate sustainability in their work. We just need to be shown how sometimes.

\bibliographystyle{elsarticle-harv} 
\bibliography{main}

\end{multicols}

\end{document}